\documentclass[letterpaper, 10 pt, conference]{util/ieeeconf}
\IEEEoverridecommandlockouts

\makeatletter
\def\endthebibliography{%
	\def\@noitemerr{\@latex@warning{Empty `thebibliography' environment}}%
	\endlist
}
\makeatother

\usepackage{amsfonts}
\usepackage{amsmath}
\usepackage{graphicx}
\usepackage{subcaption}
\usepackage[ruled]{algorithm2e}
\usepackage{verbatim}
\usepackage{amssymb}

\newtheorem{theorem}{Theorem}
\newtheorem{lemma}{Lemma}
\newtheorem{problem}{Problem}
\newtheorem{prop}{Proposition}

\newtheorem{example}{Example}

\DeclareMathOperator*{\argmax}{arg\,max}
\DeclareMathOperator*{\argmin}{arg\,min}

\newcommand{\hipi}{\overline{\pi}^a}
\newcommand{\lopi}{\underline{\pi}^a}
\newcommand{\hira}{\overline{r}^a}
\newcommand{\lora}{\underline{r}^a}

\title{Stackelberg Equilibria for Two-Player Network Routing Games \\ on Parallel Networks}
\author{David Grimsman, Jo\~{a}o P. Hespanha and Jason R. Marden %
\thanks{D. Grimsman (\texttt{davidgrimsman@ucsb.edu}), J. R. Marden (\texttt{jmarden@ece.ucsb.edu}), and J. P. Hespanha (\texttt{hespanha@ece.ucsb.edu}) are with the Department of Electrical and Computer Engineering, University of California, Santa Barbara, CA}
\thanks{This research was supported by NSF Grant \#ECCS-1638214, U.S. Office of Naval Research (ONR) grant \#N00014-17-1-2060 and ONR MURI grant No. N00014-16-1-2710.}
\thanks{© 2020. This work has been accepted to American Control Conference for publication under a Creative Commons Licence CC-BY-NC-ND}}

\begin{document}
	
\maketitle

\begin{abstract}
We consider a two-player zero-sum network routing game in which a router wants to maximize the amount of legitimate traffic that flows from a given source node to a destination node and an attacker wants to block as much legitimate traffic as possible by flooding the network with malicious traffic. We address scenarios with asymmetric information, in which the router must reveal its policy before the attacker decides how to distribute the malicious traffic among the network links, which is naturally modeled by the notion of Stackelberg equilibria. The paper focuses on parallel networks, and includes three main contributions: we show that computing the optimal attack policy against a given routing policy is an NP-hard problem; we establish conditions under which the Stackelberg equilibria lead to no regret; and we provide a metric that can be used to quantify how uncertainty about the attacker's capabilities
limits the router's performance.
\end{abstract}
\section {Introduction}

This paper addresses a network routing game between a player that
wants to route legitimate traffic from a source node to a destination
node and another player that wants to block traffic by flooding the
network with malicious traffic. We refer to these players as the
\emph{router} and the \emph{attacker}. Motivated by network security
problems, we are interested in scenarios of asymmetric information,
where the router exposes its policy to the attacker before the
attacker needs to select its policy. The problem formulation
considered here is motivated by the so-called Crossfire attack in
which an attacker persistently degrades network connectivity by
targeting a selected set of links within the network, while adjusting
to changes in routing policies \cite{kang2013crossfire}. The defense against such attacks has been the subject of recent work \cite{gkounis2014towards,aydeger2016mitigating,gkounis2016interplay,raj2018crossfire}.

The Nash equilibrium is an attractive solution concept for noncooperative
games because it leads to very strong notions of equilibria, in that
neither player regrets its choice after the outcome of the game is
revealed \cite{nash1951non}. However, such equilibria often do not exist in problems of
asymmetric information. The Stackelberg equilibrium is an alternative
solution concept where one player (the leader) must select and reveal
its policy before the other player (the follower) makes a
decision \cite{von2010market}. This type of equilibrium specifically addresses the
information asymmetry that we consider here and has been applied to
domains closely related to the problem considered in this paper,
including network routing \cite{korilis1997achieving}, scheduling \cite{roughgarden2004stackelberg}, and channel allocation for cognitive radios \cite{bloem2007stackelberg}, but also has application in supply chain and marketing channels \cite{he2007survey} among other fields. The Stackelberg equilibrium is a concept that is also well-suited for security of critical infrastructure systems \cite{brown2006defending} and has been applied to
surveillance problems that include the ARMOR program at the Los
Angeles International Airport \cite{pita2008deployed}, the IRIS program used by the US
Federal Air Marshals \cite{jain2010software}, power grid security \cite{brown2005analyzing}, and defending oil reserves \cite{brown2006defending}. These two types of equilibria have also been studied extensively for various types of security games \cite{korzhyk2011stackelberg}.

This paper includes three main contributions: 
\begin{enumerate}
	\item Theorem \ref{thm:hard} establishes that computing the attacker's optimal response to a routing policy is an NP-hard problem.
	\item Theorem \ref{thm:equilibria} determines conditions on the network under which Stackelberg equilibria lead to no-regret policies (i.e., are also Nash).
	\item Section \ref{sec:equilibria} explores how uncertainty in knowledge about the capabilities of the attacker translates into performance loss for the router. Theorem \ref{thm:twolink} provides a closed-form expression which quantifies this for a two-link network.
\end{enumerate}

We focus on a network consisting solely of $N$ parallel links that directly connect source and destination. Even within this simple set of networks, the computation of the optimal attack policy turns out to have higher complexity than one might expect. For any fixed routing policy, we show in Section \ref{sec:hard} that the computation of the ``optimal'' distribution of a fixed budget of attack traffic among the parallel links is an NP-hard problem with respect to the scaling parameter $N$. From the attacker's perspective, ``optimal'' means that the attacker can prevent as much traffic as possible from reaching the destination, by flooding network links so that legitimate traffic in excess the links' capacity is dropped.

As noted above, Nash equilibria have the desirable feature that they lead to no regret by both players, a feature that is generally not shared by Stackelberg equilibria. It turns out that in the network routing games considered here, Stackelberg equilibria only lead to no-regret (i.e., are also Nash equilibria) in the extreme cases where the attacker controls a very large or a very small amount of traffic. We show this to be true for parallel networks in Section \ref{sec:equilibria}. For these two extreme cases, we actually provide explicit formulas for the optimal Stackelberg/Nash routing policies. Not surprisingly in view of the NP-hardness result, no explicit formulas are provided for intermediate levels of attack traffic.

Motivated by the nontrivial dependence of the Stackelberg policy on the total amount of traffic $r^a$ controlled by the attacker, we also study how uncertainty in $r^a$ affects routing performance. Previous work in this area has modeled this type of uncertainty as a distribution over the possible values of $r^a$, giving rise to routing policies that give an optimal expected value on the cost function \cite{paruchuri2008playing}. However, in this work, we define a metric for the ``value of information'' about the power of the attacker that compares the amount of traffic that the attacker could block if the router knew precisely $r^a$ versus the amount of traffic it could block if the router had to select a policy without precise knowledge of $r^a$. The latter scenario generally leads to an increase in blocked traffic. We show in Section \ref{sec:value} a closed-form expression for the value of information in two-link networks.

%We show in Section \ref{sec:value} that our notion of value of information can be computed for parallel networks by solving a minimization over a convex set of routing policies of a criteria that only needs to consider a fixed number of values for the total amount of traffic $r^a$ (even though $r^a$ generally takes value in a continuum). Closed-form solutions to this optimization can be found for two-link networks.
\section{Model}

This paper focuses on a two-player network routing game where the system operator is tasked with deriving a routing policy to maximize the throughput of a given single source~/ single destination parallel network in the presence of an adversary. The network is comprised of a set of edges $E$, where each edge $e \in E$ is associated with a given capacity $c_e \geq 0$. The system operator, which we will henceforth refer to as the \emph{router}, is tasked with with designing a routing profile $f = \{f_e\}_{e \in E}$ which routes $r \geq 0$ units of traffic across this network. A feasible routing profile satisfies $\sum_{e \in E} f_e = r$ and $0 \leq f_e \leq c_e$ for all edges $e \in E$. We denote the convex set of all admissible routing profiles as ${\cal F}(c, r)$ where  $c = \{c_e\}_{e\in E}$ denotes the capacities of all edges. 

This work considers the existence of an attacker whose goal is to block as much routed traffic as possible by reducing the capacities of the edges in the network through a cross-fire style attack where the attacker can send up to $r_a \geq 0$ units of non-responsive traffic on various edges in the network.  An adversarial attack can be characterized by a routing profile $f^a = \{f_e^a\}_{e \in E}$ which satisfies $\sum_{e \in E} f_e^a = r^a$ and $0 \leq f_e^a \leq c_e$ for all edges $e \in E$.  We denote the set of all admissible adversarial attack policies as ${\cal F}^a(c, r^a)$. We will often refer to $r^a$ as the attack budget of the adversary.  Given an admissible routing profile $f \in {\cal F}(c, r)$ and an adversarial attack $f^a \in {\cal F}^a(c, r^a)$, the amount of legitimate traffic blocked on any edge $e \in E$ is defined as
\begin{equation} \label{eq:blockdef}
    B_e(f, f^a, c) := \max\left\{f_e + f_e^a - c_e, 0\right\},
\end{equation}
and the total blocked traffic in the system as $B(f, f^a, c) = \sum_{e \in E} B_e(f, f^a, c)$. Since the routing policy is non-responsive, the adversarial choice effectively reduces the capacity on each edge $e$ from $c_e$ to $c_e - f_e^a$. Lastly, we will often omit highlighting the functional dependence on the parameters $c$, $r$, and $r^a$ for brevity, e.g., express ${\cal F}^a(c, r^a)$ as merely ${\cal F}^a$, when this dependence is clear.  

One focus of this paper is to characterize different forms of equilibria in this two-player network routing game. In general, we will assume that a router is required to choose the routing strategy first and the adversary can respond accordingly.  The most natural class of equilibria that captures this phenomena is that of Stackelberg equilibria (SE), which consists of any pair of routing profiles $(f,f^a)$ such that 
\begin{eqnarray}
f &\in& \underset{\bar{f} \in {\cal F}}{\arg \inf} \ \sup_{\bar{f}^a \in {\cal F}^a} B(\bar{f}, \bar{f}^a, c), \label{eq:se_r} \\
f^a &\in& \underset{\bar{f}^a \in {\cal F}^a}{\arg \sup}  \ B(f, \bar{f}^a, c). \label{eq:se_a}
\end{eqnarray}
If $f^a$ satisfies \eqref{eq:se_a}, we refer to $f^a$ as a \emph{best response attack} to $f$. A second class of equilibria that we focus on is Nash equilibria (NE), which focuses on situations where both the router and adversary are required to select their strategy without knowledge of the other's choice.  A NE is defined as any pair of profiles $(f,f^a)$ such that 
\begin{eqnarray}
f &\in& \underset{\bar{f} \in {\cal F}}{\arg \inf} \  B(\bar{f}, {f}^a, c), \label{eq:ne_r}\\
f^a &\in& \underset{\bar{f}^a \in {\cal F}^a}{\arg \sup}  \ B(f, \bar{f}^a, c). \label{eq:ne_a}
\end{eqnarray}
We refer to $\mathcal{SE}(r, r^a)$ as the set of all SE for values $r, r^a$, and likewise $\mathcal{NE}(r, r^a)$ for NE. Note that given the definitions above, $\mathcal{NE}(r, r^a) \subseteq \mathcal{SE}(r, r^a)$. In the event where $\mathcal{NE}(r, r^a) = \mathcal{SE}(r, r^a)$, this implies that the router is not strategically disadvantaged by having to reveal its choice before the adversary selects its policy.  However, while a SE will always exist, the same does not hold true for NE. Furthermore, this paper will address how knowledge of the exact value of $r^a$ impacts the existence and efficacy of such equilibria.

\begin{example} \label{ex:5link}

\begin{figure}
	\centering
	\begin{subfigure} {0.48\textwidth}
	    \centering
	    \includegraphics[scale=0.5]{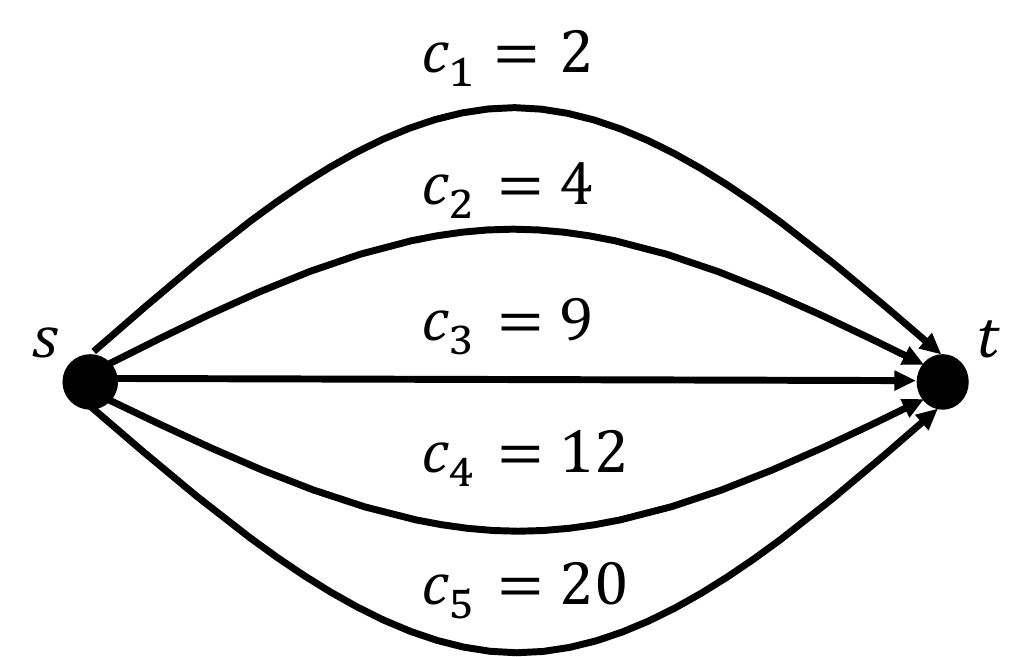}
	    \caption{An example network. Suppose that $f = \{1, 1, 5, 10, 8\}$ and $f^a = \{2, 4, 4, 4, 6\}$. Then, for instance on edge 4, since the capacity is 12, 2 units of traffic are blocked. In total we see that $B_1 = 1$, $B_2=1$, $B_3=0$, $B_4=2$, and $B_5=0$, which results in $B(f, f^a, c) = 4$.}
	    \label{fig:ex_5}
	\end{subfigure}
	\begin{subfigure} {0.48\textwidth}
	    \centering
	    \includegraphics[scale=0.39]{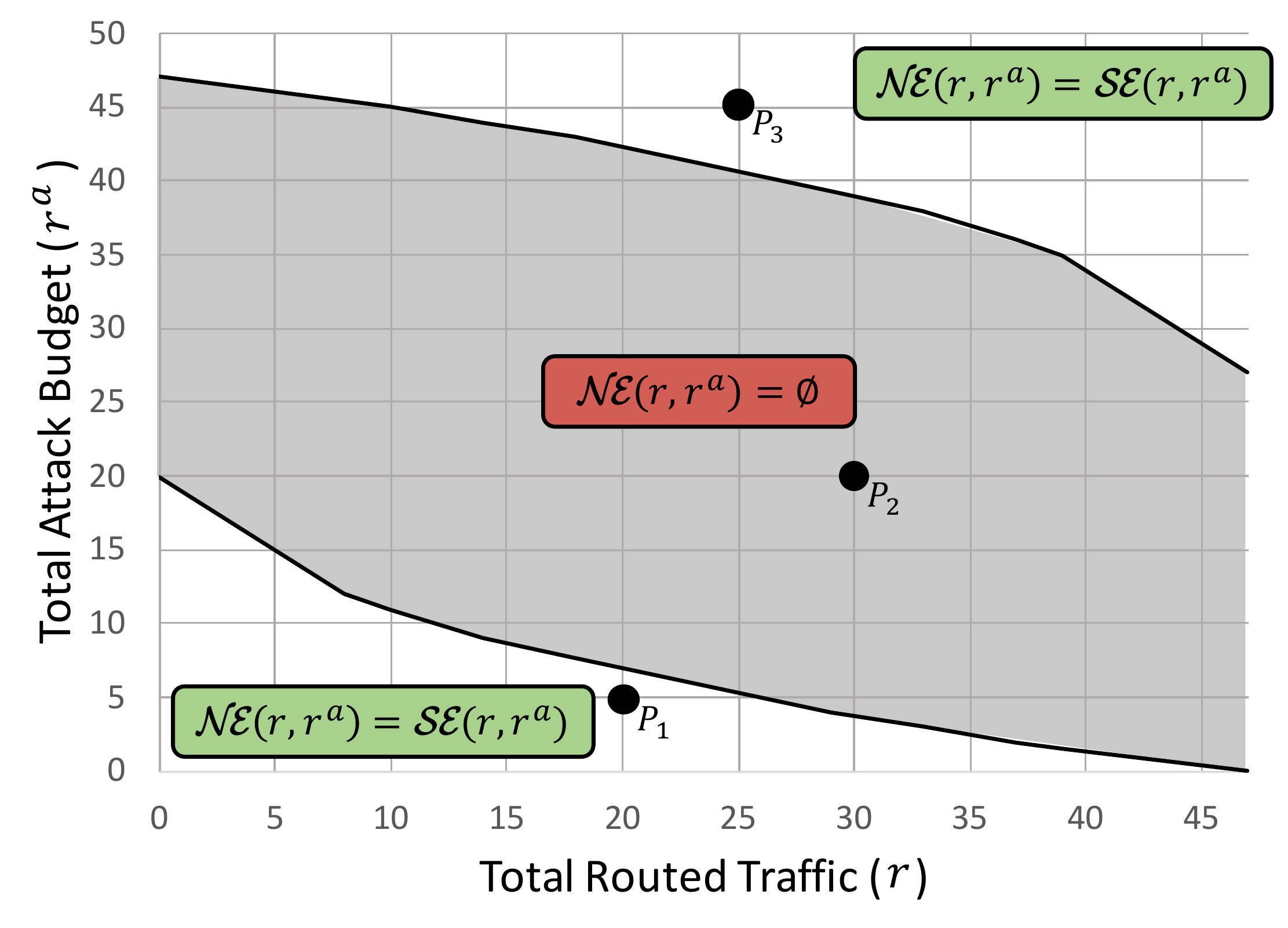}
	    \caption{This figure showcases one of the contributions of this paper: a characterization of where no NE exist (gray region) and when all SE are also NE (white regions) for the network in (a). For any values $(r, r^a)$, one of those two properties must hold. See Example \ref{ex:5link} and Theorem \ref{thm:equilibria} for more details.}
	    \label{fig:se_v_ne}
	\end{subfigure}
	\caption{An example network showcasing the model and regions of $(r, r^a)$ where NE exist.}
	\label{fig:ex_model}
\end{figure}

We begin with the following example highlighting the complexity of computing NE and SE in such a routing game.  To that end, consider the example shown in Figure \ref{fig:ex_5} with $r = 25$ and $r^a = 20$ and denote the edge set as $E = \{1, 2, \dots, 5\}$ and edge capacities as $c = \{2, 4, 9, 12, 20\}$. Given a routing profile $f = \{1, 1, 5, 10, 8\}$ and an attack profile $f^a = \{2, 4, 4, 4, 6\}$, it follows from \eqref{eq:blockdef} that the traffic blocked on each edge is 1, 1, 0, 2 and 0, respectively.  Note that these strategy profiles $(f,f^a)$ neither capture a NE or SE as there are numerous adversarial strategies that could increase the total blocked traffic given the routing profile $f$, e.g.,  $\bar{f}^a=\{0, 0, 8, 12, 0\}$.

The plot in Figure \ref{fig:se_v_ne} highlights the distinction between NE and SE for the considered routing problem for all pairs $(r,r^a)$ satisfying $47 \geq r,r^a \geq 0$. For instance, when $r=20$ and $r^a = 5$ (see point $P_1$ in Figure \ref{fig:se_v_ne}), any SE is also a NE. One such routing profile is $f = \{0, 0, 0, 5, 15\}$, as this does not allow the attacker to block any traffic. When $r = 25$ and $r^a = 45$ (point $P_3$), we see a similar phenomenon, where the attacker has much more power. In fact, observe that the routing profile $f = \{2, 4, 6.\bar{3}, 6.\bar{3}, 6.\bar{3}\}$ and attack profile $f^a = \{0, 4, 9, 12, 20\}$ constitute both a SE and NE. The router is able to design a policy such that the attacker can only block $\sum_{e \in E}c_e - r^a$ traffic, the best the router can achieve given $r^a$. Thus the router has no incentive to deviate, and clearly the attacker cannot. Lastly, when $r = 30$ and $r^a = 20$ (point $P_2$) we begin to notice a discrepancy between NE and SE in the sense that given any profiles $(f,f^a)$, if \eqref{eq:ne_a} is satisfied then \eqref{eq:ne_r} is not satisfied. For example, consider the profiles $f = \{1.4, 4, 6.4, 6.4, 11.8\}$ and $f^a = \{0, 0, 0, 0, 20\}$ and note that $f^a$ satisfies \eqref{eq:ne_a}. If the attacker implements this policy, then $(f, f^a)$ is not a NE, since the router would benefit unilaterally by moving some traffic from edge 5 to another unblocked edge. The forthcoming Theorem \ref{thm:equilibria} provides the characterization shown in Figure~\ref{fig:se_v_ne}.

\end{example}
\section{Problem Hardness} \label{sec:hard}

In this section, we show that finding the best response attack policy in \eqref{eq:se_a} is NP-Hard. We formally define it as follows:

\begin{problem} \label{prob:opt_a}
	Given a parallel network with edges $E$, corresponding capacities $c$, a routing policy $f$, and attack power $r^a$, find $f^a$ which satisfies \eqref{eq:se_a}, i.e., a best response attack policy.
\end{problem}

Note that an instance of the problem can be defined by $(E, c, f, r^a)$, and we show how the complexity of the problem scales with the number of edges in the parallel network.

\begin{theorem} \label{thm:hard}
   Problem \ref{prob:opt_a} in NP-Hard on the scaling variable $|E|$.
\end{theorem}

The theorem is proved by reducing the 0-1 Knapsack Problem (KP), a known NP-Hard problem, to Problem \ref{prob:opt_a}. We do this by showing that if all $f_e$ are ``sufficiently small", then any best response attack must either block all traffic on an edge or block none of it. Thus finding the best response attack is simply finding the set of edges to fully block, corresponding to the discrete nature of the items in the 0-1 KP. This implies any method for solving these instances of Problem \ref{prob:opt_a} will also solve the 0-1 KP.

The following lemma defines ``sufficiently small" in this context:
\begin{lemma} \label{lem:01block}
    Consider an instance of Problem \ref{prob:opt_a} $(E, c, f, r^a)$, where
    \begin{equation}
        f_e < \min_{E' \subseteq E: r^a - C(E') >0} r^a - C(E'), \label{eq:felt}
    \end{equation}
    for some $e \in E$. Then $B_e(f, f^a) \in \{0, f_e\}$ for any $f^a$ which is a solution to Problem \ref{prob:opt_a}.
    
    \begin{proof}
        We prove the contrapositive statement. Let $e$ be such that $B_e(f, f^a) \notin \{0, f_e\}$. Define $E^{\rm block} := \{e' \in E: f_{e'} + f^a_{e'} > c_e\}$, and observe by definition that $e \in E^{\rm block}$. Then it must be true that $f_e > r^a - C(E^{\rm block} \setminus \{e\}) > 0$, otherwise the attacker could block more routed traffic by redistributing as much attack traffic as possible from $e$ to the other edges in $E^{\rm block}$. Therefore, \eqref{eq:felt} must be false.
    \end{proof}
\end{lemma}

Given this, we proceed with the proof of Theorem \ref{thm:hard}. The 0-1 KP can be defined as follows: assume we have $n$ items, where each item $e$ has a cost $w_e$ and a value $v_e$. Given a total cost constraint $W$, find the combination of items with maximum total which does not exceed $W$. More formally stated, determine
\begin{equation} \label{eq:kp}
	\begin{aligned}
	& \underset{x}{\text{maximize}} & & \sum_e v_ex_e \\
	& \text{subject to}
	& & \sum_i x_e, w_e \leq W,
	& & x_e \in \{0, 1\},
	\end{aligned}
\end{equation}
where $x:=[x_e]$. This problem is known to be NP-Hard in the number of items~\cite{kellerer2004knapsack}.

Mapping a 0-1 KP to Problem \ref{prob:opt_a} can be done with the following method: let every item be mapped to an edge in a parallel network, $r^a = W$, $c_e = w_e$, and $f_e = \varepsilon v_e$, where $\varepsilon>0$ satisfies
\begin{equation}
	\varepsilon v_e < \min_{E' \subseteq E: r^a - C(E') >0} r^a - C(E'),
\end{equation}
for all $e \in E$. By Lemma \ref{lem:01block}, we know that any solution to this subset of instances of Problem \ref{prob:opt_a} has the property that every edge will either have all routed traffic blocked or none. Therefore, the problem can be reformulated as
\begin{equation} \label{eq:kp_opt_a}
\begin{aligned}
& \underset{x}{\text{maximize}} & & \sum_e f_e x_e \\
& \text{subject to}
& & \sum_i x_e c_e \leq r^a,
& & x_e \in \{0, 1\}.
\end{aligned}
\end{equation}
This problem yields an equivalent solution to that in \eqref{eq:kp}, since the constraints are the same, and each objective function is a scaled version of the other. Thus solving this instance of Problem \ref{prob:opt_a} will also solve 0-1 KP and shows that Problem \ref{prob:opt_a} is NP-Hard. $\blacksquare$

\begin{comment}

\begin{problem} \label{prob:opt_a}
	Given a parallel network with edges $E$, corresponding capacities $c$, a total amount of traffic $r$ that must be routed through the graph, a routing policy $f$, and an attack power $r^a$, determine
	\begin{equation}
	    f^{a*}(c, r, f, r^a) \in \argmax_{f^a \in \mathcal{F}^a} B(f, f^a, c)
	\end{equation}
\end{problem}

An instance of Problem \ref{prob:opt_a} is defined by the tuple $(E, c, r, f, r^a)$ and we are interested in how the complexity scales with the number of edges in the network.

\begin{prop} \label{prop:hard}
	Problem \ref{prop:hard} is NP-Hard on the scaling variable $|E|$.
\end{prop}

The full proof for Proposition \ref{prop:hard} is found in Appendix\ref{sec:proof_hardness}, but essentially, the 0-1 Knapsack Problem (a known NP-Hard problem) can be reduced to the optimal attack problem. Then any solver which can solve Problem \ref{prob:opt_a} can also solve 0-1 Knapsack Problem, completing the proof. 

\end{comment}
\section{Equilibria} \label{sec:equilibria}

In this section, we present results that describe precisely the relationship between SE and NE in our model. For some $E' \subseteq E$, we denote it's total capacity as $C(E') := \sum_{e \in E'} c_e$. 

\begin{theorem} \label{thm:equilibria}
	Consider a parallel network with capacities $c$, routing demand $r$, and adversarial routing power $r^a$. The set of Nash Equilibria $\mathcal{NE}(r, r^a)$ is nonempty and 
	$\mathcal{NE}(r, r^a) = \mathcal{SE}(r, r^a)$ if and only if one of the following is satisfied: \footnote{While finding the maxima in \eqref{eq:low_a} and \eqref{eq:high_a} may appear to be computationally intractable given the number of edges in the network, it is true that the maximizing $E'$ for both \eqref{eq:low_a} and \eqref{eq:high_a} is of the form \{1, 2, \dots, k \}, where the edges are ordered starting with highest capacity to the lowest. Therefore, finding either maxima is equivalent to finding the best value of $k$, which can be completed in linear time.}
	\begin{align}
	    &r^a \leq \max_{E' \subseteq E}\frac{C(E') - r}{|E'|} \label{eq:low_a}  \\ 
		&r^a \geq C(E) - \max_{E' \subseteq E}\frac{r - C(E\setminus E')}{|E'|}.  \label{eq:high_a}
    \end{align}
    
    \begin{proof}
        Note that since $B(f, f^a) = \sum_e \max\{ f_e + f^a_e - c_e, 0\}$, then a lower bound on $B(f, f^a)$ is
            \begin{equation} \label{eq:block_lb}
                B(f, f^a) \ge r + r^a - C(E).
            \end{equation}
        We now begin with a few observations about router best responses:
        \begin{enumerate}
            \item For a policy pair $(f, f^a)$, if $f_e + f^a_e \le c_e$ for all $e$, then $B(f, f^a) = 0$, and the router has no incentive to deviate. If $f^a$ satisfies \eqref{eq:ne_a}, then $(f, f^a)$ is both a SE and a NE.
            \item For a policy pair $(f, f^a)$, if $f_e + f^a_e \ge c_e$ for all $e$, then $B(f, f^a) = r + r^a - C(E)$, the lower bound in \eqref{eq:block_lb}. Thus the router has no incentive to deviate. If $f^a$ also satisfies \eqref{eq:ne_a}, then $(f, f^a)$ is both a SE and a NE.
            \item For a policy pair $(f, f^a)$, if there exist $e, e' \in E$ such that $f_e + f^a_e > c_e$ and $f_{e'} + f^a_{e'} < c_{e'}$, then \eqref{eq:ne_r} is not satisfied. Therefore, $(f, f^a)$ is not a NE.
        \end{enumerate}
        We now proceed with proving Theorem \ref{thm:equilibria}. To that end, consider the routing policy $f^{\rm lo}$, where
        \begin{equation} \label{eq:low_a_rt}
    		f^{\rm lo}_e := \max\left\{c_e - \max_{E' \subseteq E}\frac{C(E') - r}{|E'|}, 0\right\}.
    	\end{equation}
    	We first show that $f^{\rm lo}$ is feasible. Let $E^*$ be the largest set in $\argmax_{E' \subseteq E}(C(E') - r)/|E'|$. Then $e \in E^*$ if and only if
    	\begin{align}
    	    \frac{C(E^*) - r}{|E^*|} \geq & \frac{C(E^* \setminus \{e\}) - r}{|E^* \setminus \{e\}|}. \\
    	    =& \frac{C(E^*) - c_e - r}{|E^*| - 1} \implies \\
            %(|E^*| - 1)(C(E^*) - r) \geq & |E^*|(C(E^*) - c_e - r) \iff \\
            %(|E^*| - 1)(C(E^*) - c_e - r)& + (|E^*| - 1)c_e \geq \nonumber \\
            %(|E^*| - 1)(C(E^*) - c_e - r)& + C(E^*) - c_e - r \iff \\
            %(|E^*| - 1)c_e \geq & C(E^*) - c_e - r \iff \\
            %|E^*|c_e \geq & C(E^*) - r \iff \\
            c_e \geq & \frac{C(E^*) - r}{|E^*|},
    	\end{align}
    	where we define $r/0 = \infty$. Since this is true, it follows that
    	\begin{align}
            \sum_{e \in E} f^{\rm lo}_e =& \sum_{e \in E^*} f^{\rm lo}_e, \\
            =& \sum_{e \in E^*} c_e - \frac{C(E^*) - r}{|E^*|}, \\
            =& C(E^*) - |E^*| \frac{C(E^*) - r}{|E^*|} = r,
        \end{align}
    	thus $f^{\rm lo}$ is feasible. 
    	
    	Note that if $r^a$ satisfies \eqref{eq:low_a}, then for any allowable attack $f^a$, $f_e^{\rm lo} + f^a_e \le c_e$ for all $e$. Hence by observation 1, $(f, f^a)$ is a SE and a NE. Since $B(f, f^a) = 0$ must hold for any SE, we conclude that $\mathcal{SE}(r, r^a) = \mathcal{NE}(r, r^a)$.
    	
    	We now turn our attention the case in \eqref{eq:high_a}. To this end, consider the routing policy $f^{\rm hi}$, where
    	\begin{equation}
    	    f^{\rm hi}_e := \min\left\{c_e, \max_{E' \subseteq E}\frac{r - C(E\setminus E')}{|E'|}\right\}.
    	\end{equation}
    	
    	\begin{comment}
    	We first show that $f^{\rm hi}$ is feasible. Let $E^*$ be the largest set in $\argmax_{E' \subseteq E} (r - C(E \setminus E'))/|E'|$. Then $e \in E^*$ if and only if
    	\begin{align}
            \frac{r - C(E \setminus E^*)}{|E^*|} \geq & \frac{r - C(E \setminus (E^* \setminus \{e\}))}{|E^* \setminus \{e\}|}, \\
            =& \frac{r - C(E \setminus E^*) - c_e}{|E^*| - 1} \implies \\
            %(|E^*| - 1)(r - C(E \setminus E^*)) \geq & |E^*|(r - C(E \setminus E^*) - c_e) \iff \\
            %(|E^*| - 1)(r - C(E\setminus E^*) - &c_e) + (|E^*| - 1)c_e \geq \nonumber \\
            %(|E^*| - 1)(r - C(E\setminus E^*) - &c_e) + r - C(E \setminus E^*) - c_e \iff \\
            %(|E^*| - 1)c_e \geq & r - C(E \setminus E^*) - c_e \iff \\
            %|E^*| c_e \geq & r - C(E \setminus E^*) \iff \\
            c_e \geq & \frac{r - C(E \setminus E^*)}{|E^*|},
        \end{align}
    	where again we define $r/0 = \infty$. Since this is true, it follows that
    	\begin{align}
            \sum_{e \in E} f^{\rm hi}_e =& \sum_{e \notin E^*} c_e + \sum_{e \in E^*} \frac{r - C(E \setminus E^*)}{|E^*|}, \\
            =& C(E \setminus E^*) + |E^*|\frac{r - C(E \setminus E^*)}{|E^*|}, \\
            =& r, 
        \end{align}
    	thus $f^{\rm hi}$ is feasible.
    	\end{comment}
    	
    	This policy is feasible, which can be shown using a similar argument as that given above for the feasibility of $f^{\rm lo}$. If $r^a$ satisfies \eqref{eq:high_a}, then for any allowable attack $f^a$, $f_e^{\rm hi} + f^a_e \geq c_e$ for all $e$. By observation 2, $(f^{\rm hi}, f^a)$ is a SE and a NE. Since $B(f, f^a) = r + r^a - C(E)$ for any SE, we conclude that $\mathcal{NE}(r, r^a) = \mathcal{SE}(r, r^a)$.
    	
    	Suppose that $r^a$ does not satisfy \eqref{eq:low_a}. Let $f \in \mathcal{F}(c, r)$ and denote $E^{\rm flow} = \{e: f_e > 0\}$. Then
    	\begin{align}
    	    r^a >& \max_{E' \subseteq E} \frac{C(E') - r}{|E'|} \ge \frac{C(E^{\rm full}) - r}{|E^{\rm full}|}
    	    %=& \frac{\sum_{e \in P}c_e - \sum_{e \in P} f_e}{|P|}, \\
    	    %=& \frac{\sum_{e \in P}(c_e - f_e)}{|P|}, \\
    	    \ge  \min_{e \in E^{\rm full}} c_e - f_e. \label{eq:mincefe}
    	\end{align}
    	Here we have omitted some of the algebra to allow for space constraints. If $e'$ minimizes the expression in the righthand side of \eqref{eq:mincefe}, then there exists an $f^a$ such that $f_{e'} + f^a_{e'} > c_{e'}$. Since $B(f, f^a) > 0$, it must be true that for any SE $(f, f^a)$, there must be an edge $e$ where $f_e + f^a_e > c_e$.
    	
    	Suppose that $r^a$ does not satisfy \eqref{eq:high_a}. Let $f \in \mathcal{F}(c, r)$ and denote $E^{\rm part} = \{e: f_e < c_e\}$. Then
    	\begin{align}
    	    r^a < & C(E) - \max_{E' \subseteq E} \frac{r - C(E \subseteq E')}{|E'|}
    	    %\le &  C(E) - \frac{r - C(E \setminus E^{\rm part})}{|E^{\rm part}|}, \\
    	    %=& C(E) - \frac{\sum_{e} f_e - \sum_{e \notin P} c_e}{|P|}, \\
    	    %=& C(E) - \frac{\sum_{e \in P} f_e}{|P|}, \\
    	    \le C(E) - \min_{e \in E^{\rm part}} f_e, \label{eq:minfe}
    	\end{align}
    	where again we have omitted the algebra for the sake of space. If $e'$ minimizes the rightmost expression in \eqref{eq:minfe}, then there must exist an attack policy $f^a$ where $f_{e'} + f^a_{e'} < c_{e'}$. Since $B(f, f^a) > r + r^a + c$, it must be true that for any SE $(f, f^a)$, there must be an edge $e$ where $f_e + f^a_e < c_e$. Therefore, by observation 3 we conclude that when $r^a$ satisfies neither \eqref{eq:low_a} nor \eqref{eq:high_a}, no NE can exist.
    \end{proof}
\end{theorem}

Refer again to the network in Figure \ref{fig:ex_model}. At point $P_1$, $r=20$ and $r^a=5$. Here we calculate $\max_{E' \subseteq E}(C(E') - r)/|E'| = 7$, which means that $r^a$ satisfies \eqref{eq:low_a}. Thus the router can use the policy $f^{\rm lo } = \{0, 0, 2, 5, 13\}$ to ensure that the attacker cannot block any traffic. By Theorem \ref{thm:equilibria}, this also implies that $(f, f^a)$ is both a SE and a NE for any $f^a \in \mathcal{F}^a(c, r)$. At point $P_3$, $r=25$ and $r^a = 45$. Here we calculate $C(E) - \max_{E' \subseteq E}(r - C(E\setminus E'))/|E'| = 40.\bar{6}$, which means that $r^a$ satisfies \eqref{eq:high_a}. Thus the router can use the policy $f = \{2, 4, 6.\bar{3}, 6.\bar{3}, 6.\bar{3}\}$, and from Theorem \ref{thm:equilibria}, $(f, f^a)$ is a NE and SE for any $f^a \in \mathcal{F}^a(c, r)$. At point $P_2$, $r=30$ and $r^a=20$. We calculate that $\max_{E' \subseteq E} (C(E') - r)/|E'| = 3.75$ and $C(E) - \max_{E' \subseteq E}(r - C(E \setminus E'))/|E'| = 59$, therefore $r^a$ does not satisfy \eqref{eq:low_a} or \eqref{eq:high_a}. By Theorem \ref{thm:equilibria}, we know that no NE can exist at this point.

\begin{figure}
    \centering
    \includegraphics[scale=0.42]{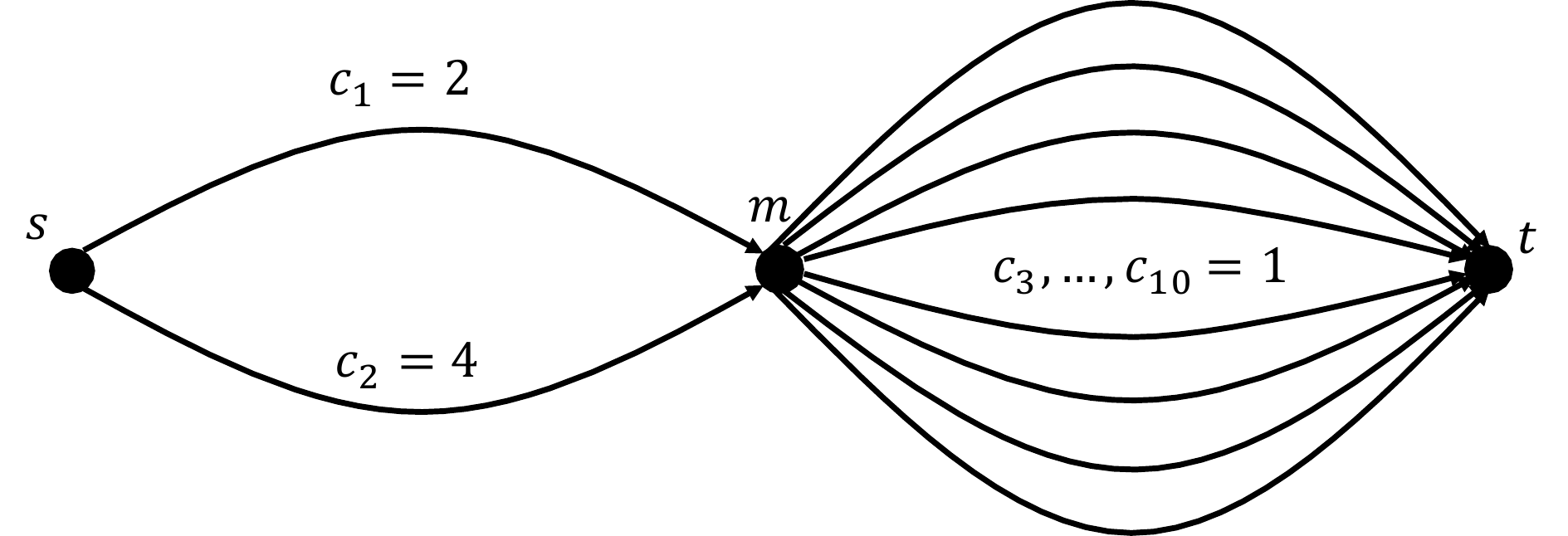}
    \caption{Two parallel networks in series. We use this example to illustrate the complexities for finding SE and NE in more general networks than just parallel. For instance, one cannot simply decompose the optimal attack problem into either attacking the set of edges between $s$ and $m$, and attacking the edges between $m$ and $t$. Even if we limited our scope to such attacks, which set of edges to attack depends on the value of $r^a$, not merely on $f$ and $c$. See Example \ref{ex:nonparallel} for more details.}
    \label{fig:ex_hard}
\end{figure}

\begin{example} \label{ex:nonparallel}
Consider now the example in Figure \ref{fig:ex_hard}, a graph where two parallel networks are connected in series. We present this as a simple example to showcase the complexities that arise when studying the SE of non-parallel networks. For more complex networks, one might think that finding a best response attack could be limited to attacking a minimal cut-set in the network. However, even in this very simple example, we show that this isn't the case, and in fact, a best response attack will often incorporate edges of multiple cut-sets in the network. Thus investigating parallel networks in this paper gives a natural simplification of the problem in order to address the questions of interest.

In Figure \ref{fig:ex_hard}, denote $E_{sm}$ as the cut-set of edges between $s$ and $m$ and $E_{mt}$ as the cut-set of edges between $m$ and $t$. Observe that regardless of the attacker's capability, there always exists a SE route where all edges in $E_{mt}$ have the same amount of traffic routed on them. We assume in the following cases that the router always uses such a policy, and therefore, we need only focus on the routing strategy across $E_{sm}$. 

Let $r = 2$ and $r^a = 5$. If the attacker restricts its attacks to a single cut-set $E_{sm}$ or $E_{mt}$, then the router can choose its policy accordingly, for instance $f_e = 1$ for $e \in E_{sm}$, and $f_e = 0.25$ for $e \in E_{mt}$. Note that across each cut-set, this route satisfies \eqref{eq:se_r}. Attacking only $E_{sm}$, the attacker can block 1 unit of traffic, but attacking only $E_{mt}$, the attacker can block 1.25 units of traffic. This may seem unintuitive, since the total capacity of $E_{sm}$ is less than that of $E_{mt}$. Furthermore, the best response for the attacker is to block some traffic on $E_{sm}$ and some on $E_{mt}$. For instance, the attacker could block the 1 unit of traffic on edge 1, and then block all traffic on 3 of the edges in $E_{mt}$. Assuming that the router evenly distributes the remaining 1 unit of routed traffic that arrives at node $m$, this attack would block 1.375 units of traffic. Therefore, solving for a SE must include all attacks across multiple cut-sets.

Given these complexities with even very simple non-parallel networks, the characterizations of SE and NE in Theorem \ref{thm:equilibria} only apply to parallel networks. While this class of networks is sufficiently rich to ask the questions and showcase the phenomena that are relevant to this work, future work can ask similar questions in a broader setting.
\end{example}
\section{The Value of Information} \label{sec:value}

In this section, we present preliminary results about the value to the router of knowing information about the attack power $r^a$. In order to do this, we introduce some notation. We define
\begin{equation}
    B^*(f, r^a) := B(f, f^a),
\end{equation}
where $f^a \in \mathcal{F}^a(r^a)$ satisfies \eqref{eq:se_a}. In other words, $B^*(f, r^a)$ measures how much traffic is blocked in the attacker's best response to $f$, given $r^a$. We also define
\begin{equation}
    B^{\rm SE}(r, r^a) := B(f, f^a),
\end{equation}
where $(f, f^a)\in \mathcal{F}(r) \times \mathcal{F}^a(r^a)$ is a SE. Recall that for the pair $(r, r^a)$ the same amount of traffic will be blocked by any SE $(f, f^a)$.

As an example of both these functions, consider the plot in Figure \ref{fig:parallel_blocked} for a three-link parallel network where $c = \{2, 3, 5\}$ and $r=5$. For the fixed route $f = \{0.5, 2, 3.5\}$, the gray line represents how $B^*(f, r^a)$ changes as a function of $r^a$. Likewise, the orange line showcases $B^{\rm SE}(r, r^a)$ as a function of $r^a$. Observe that $B^*(f, r^a) \geq B^{\rm SE}(r, r^a)$ for all values of $r^a$.

\subsection{Limited information}

We limit the router's knowledge of $r^a$ by stating that the router only knows that $r^a$ is in some interval $\pi^a=[\underline{\pi}^a, \overline{\pi}^a]$. In light of this uncertainty, if the router chooses policy $f$, then we can define the risk of $f$ on interval $\pi^a$ as
\begin{equation} \label{eq:riskdef}
    R(f, \pi^a) := \max_{r^a \in \pi^a} \left( B^*(f, r^a) - B^{\rm SE}(r, r^a) \right).
\end{equation}
Intuitively, the value $B^*(f, r^a) - B^{\rm SE}(r, r^a)$ represents how much more traffic the attacker is able to block because the router chose policy $f$ instead of a SE policy for that value of $r^a$. Thus the risk $R(f, \pi^a)$ is the maximum such value across all $r^a \in \pi^a$. In other words, this measurement of risk shows, in the worst case, the advantage that the attacker gains by the router not knowing the true value of $r^a$.

\begin{figure}
	\centering
	\includegraphics[scale=0.45]{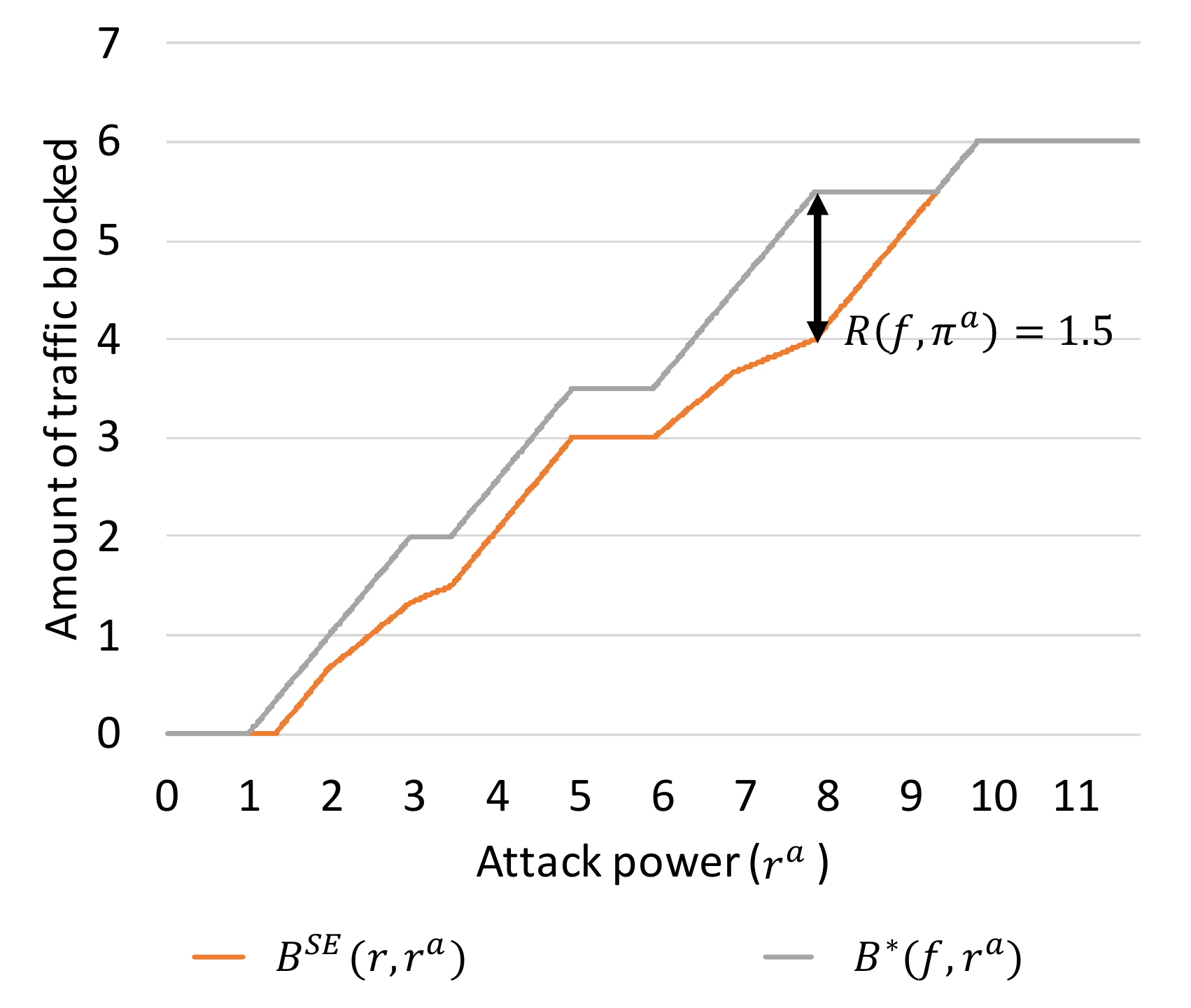}
	\caption{A plot showing the amount of traffic blocked by an optimal attack for a SE routing policy (orange) versus the traffic blocked when the router selects specific routing policy $f$ (regardless of the value of $r^a$). Here $c = \{2, 3, 5\}$, and $r = 5$. The fixed policy represented by the gray line is $f = \{0.5, 2, 3.5\}$. We show the values of the risk $R(f, \pi^a)$ for $\pi^a=[0, 10]$.}
	\label{fig:parallel_blocked}
\end{figure}

As an example, consider again the plot in Figure \ref{fig:parallel_blocked}. If we assume that the router has no knowledge of $r^a$ (i.e., $\pi^a = [0, 10]$), then the risk associated with the route $f = \{0.5, 2, 3.5\}$ is the maximum difference between the gray and orange lines, which is achieved at $r^a = 8$. Therefore, in this case we see that $R(f, \pi^a) = 1.5$.

It turns out that the maximization in \eqref{eq:riskdef} can be restricted to a finite set of points in $\pi^a$.

\begin{lemma} \label{lem:parallel}
	For a parallel network,
	\begin{equation}
		R(f, \pi^a) = \max_{r^a \in (\alpha \cap \pi^a) \cup \{\lopi, \hipi\}} B^*(f, r^a) -  B^{\rm SE}(r^a), \label{eq:mininp}
	\end{equation}
	where $\alpha$ is the finite set $\{r^a: \exists E' \subseteq E \text{ where } r^a = C(E')\}$, which has at most $2^{|E|}$ elements.
\end{lemma}

The full proof is given in Appendix\ref{app:parallelproof}, however here we provide some intuition: consider the plot in Figure \ref{fig:parallel_blocked}. The orange line, $B^{\rm SE}(r, r^a)$, is piecewise linear, with no line slope being greater than 1. The grey line, $B^*(f, r^a)$, is also a piecewise linear function, with lines slopes either 0 or 1. The value of the risk $R(f, \pi^a)$ is incurred at $r^a = C(\{2, 3\}) = 8$, where the attacker's best response against $f$ is to fully block edges 2 and 3. Because the two lines are piecewise linear, the largest distance must take place at one of the points of discontinuity for the gray line inside the interval $\pi^a$.

Finally, we define the \emph{value of information} to the router for an interval $\pi^a$ as the minimum amount of risk that can be incurred for any routing policy. More formally stated,
\begin{align} \label{eq:valdef}
    V(\pi^a) :=& \min_{f \in \mathcal{F}} R(f, \pi^a) \\
    =& \min_{f \in \mathcal{F}} \max_{r^a \in \mathcal \pi^a} \left(B^*(f, r^a) -  B^{\rm SE}(r, r^a) \right)
\end{align}
We also denote the routing policy which minimizes \eqref{eq:valdef} by $f^\pi$. This value of information is meant to reflect how valuable (i.e., how much less traffic would be blocked) if the router knew the exact value of $r^a$. For instance, if $V(\pi^a) = 0$, then there exists a route which satisfies \eqref{eq:se_r} for any value of $r^a \in \pi^a$, thus the router does not need to know the exact value. However, when $V(\pi^a)$ is high, knowing $r^a$ would allow the router to ensure that less traffic is blocked. Figure \ref{fig:two_link_info} shows $V(\pi^a)$ and $f^\pi$ for a two-link network.

\subsection{The Value of Information in Two-Link Networks}

Lemma \ref{lem:parallel} provides a numerical procedure to compute the risk for a routing policy $f$ against an attack power interval $\pi^a$ for general parallel networks. For two-link networks, this means there exists a closed-form solution for $R(f, \pi^a)$ and subsequently $V(\pi^a)$.

\begin{figure}
	\centering
	\includegraphics[scale=0.45]{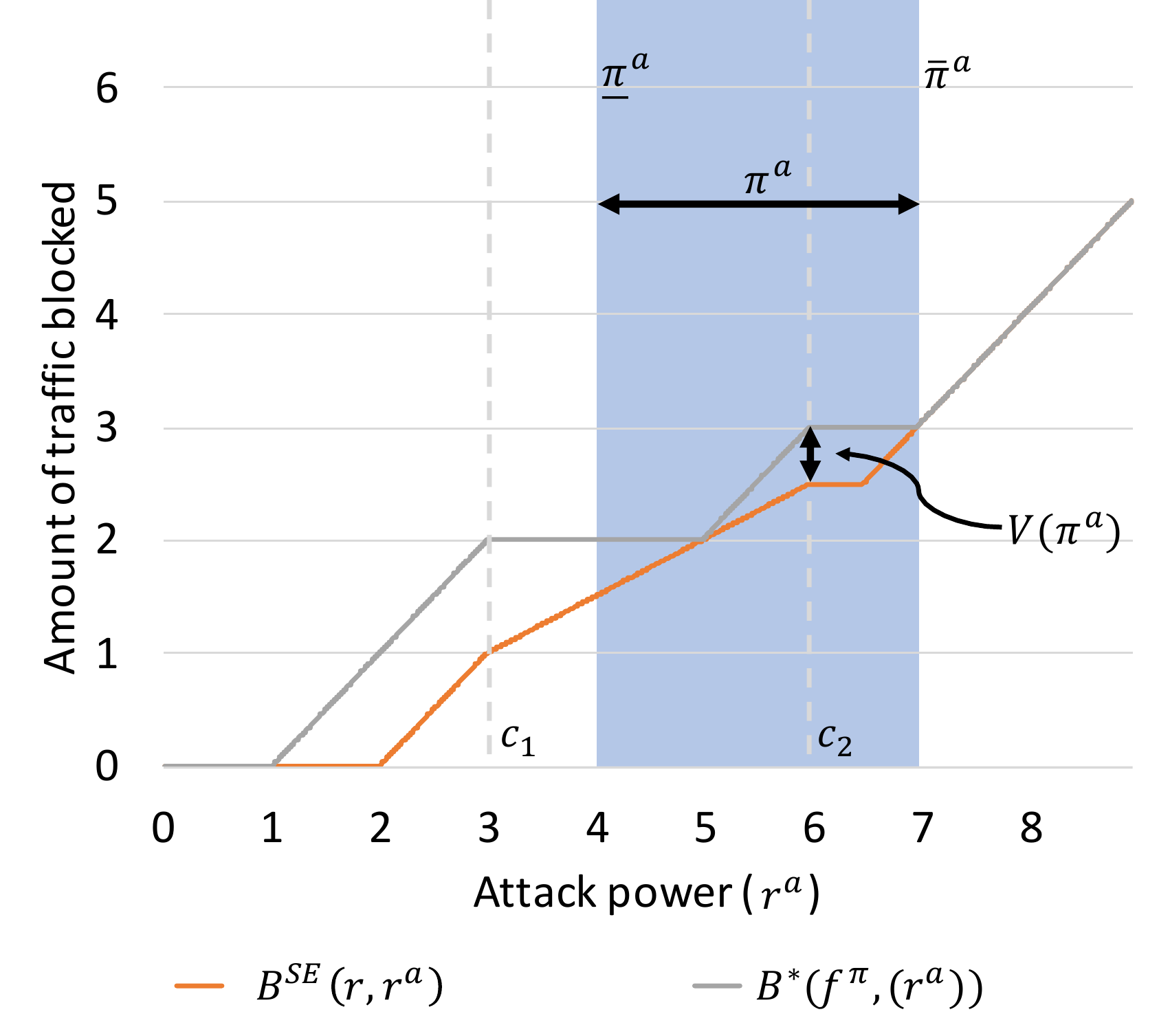}
	\caption{A plot with an example two-link parallel network that shows a graphical interpretation for $V(\pi^a)$. The edge capacities in the network are $\{3, 6\}$, and $r=5$. The orange line represents how much traffic is blocked at the SE for each value of $r^a$, and the gray line is how much is blocked by a best response attack against $f^\pi = \{2, 3\}$ for each value of $r^a$. The interval $\pi^a = [4, 7]$ is the blue shaded region. The value of information $V(\pi^a)$ is the maximum difference between the two lines within the blue region.}
	\label{fig:two_link_info}
\end{figure}

\begin{theorem} \label{thm:twolink}
 	Consider a two-link parallel network, where $c_1 \le c_2$. Suppose that the router only knows that $r^a \in \pi^a = [\lopi, \hipi]$. Then the value of information is
 	\begin{equation}
	 	V(\pi^r) = \left\{ \begin{array}{ll}
	 	    0 \text{, if } \pi^a \cap [c_1, c_2] \neq \emptyset \\
	 	    \frac{1}{4}(\min\{\hipi, c_2\} - \max\{\lopi, c_1\}) \text{, otherwise.}
	 	\end{array} \right.
 	\end{equation} 
\end{theorem}

Before proving the theorem, we first give an example to provide some intuition. Consider the plot in Figure \ref{fig:two_link_info}. In this network, $c = \{3, 6\}$, and $r=5$. If the router knows the exact value of $r^a$, it can choose a SE routing policy, which will make the difference between the lines 0 at that value of $r^a$. If we assume that the router only knows that $r^a \in \pi^a=[4, 7]$, then it must choose a policy to mitigate the risk associated with that loss of information. In this scenario, the router's best option is to use $f^\pi = \{2, 3\}$ (gray line), which minimizes the maximum difference between the two lines on $\pi^a$. The value of the routing knowing $r^a$ is then this minimum maximum difference, i.e., $V(\pi^a) = 0.5$. 

To prove Theorem \ref{thm:twolink}, we first show that we need only consider two attacks as best response.

\begin{lemma} \label{lem:2attacks}
    Consider a two-link network. For any $f$,
    \begin{equation} \label{eq:maxmax}
        B^*(f, r^a) = \max_{f^a \in \{f^{a1}(r^a), f^{a2}(r^a)\}} B(f, f^a),
    \end{equation}
    where
    \begin{align}
        &f^{a1}(r^a) := \{ \min\{r^a, c_1\}, \max\{r^a - c_1, 0\} \}, \\
        &f^{a2}(r^a) := \{ \max\{r^a - c_2, 0\}, \min\{r^a, c_2\} \}.
    \end{align}
    In other words, there always exists a best response attack policy where either (1) the attacker puts as much attack traffic as possible on edge 1 and the reminder on edge 2 (i.e., $f^{a1}(r^a)$); or (2) vice versa (i.e., $f^{a2}(r^a)$).
    
    \begin{proof}
        Let $f^a$ be a best response attack policy to $f$. If $B(f, f^a) = 0$, then the lemma is trivially true. Therefore, let $e$ be an edge where $B_e(f, f^a) > 0$, then one can create a new attack policy $\hat{f}^a$ by redistributing as much attack traffic as possible from the other edge $e'$ to $e$. Let this amount be $\delta$, so $\hat{f}^a_e = f^a_e + \delta$. Then $B_e(f, \hat{f}^a) = B_e(f, f^a) + \delta$ and $B_{e'}(f, \hat{f}^a) \geq B_{e'}(f, f^a) - \delta$. This implies $B(f, \hat{f}^a) \geq B(f, f^a)$, which is at equality since $f^a$ is a best response. Since $\hat{f}^a \in \{f^{a1}, f^{a2}\}$, we conclude the proof.
    \end{proof}
\end{lemma}

Lemma \ref{lem:2attacks} allows us to only consider two attack policies when solving for the best response, but it also gives us a simple way to solve for a SE. In the two-link case, $f$ is a SE routing policy if
\begin{equation} \label{eq:se_r_simp}
    f \in \argmin_{f \in \mathcal{F}} \max \{B(f, f^{a1}(r^a)), B(f, f^{a2}(r^a))\}.
\end{equation}
Observe that if $B(f, f^{a1}(r^a)) = B(f,  f^{a2}(r^a))$ then $f$ satisfies \eqref{eq:se_r_simp}, since moving traffic between the edges can only increase $B(f, f^{a1}(r^a))$ or $B(f, f^{a2}(r^a))$. We will leverage this observation to find $B^{\rm SE}(r, r^a)$ in the following proof.

Now we prove Theorem \ref{thm:twolink}, beginning with the case when $\pi^a \cap [c_1, c_2] = \emptyset$. First let $r^a < c_2$, and denote $g$ as the value of the maximization in \eqref{eq:low_a}. When $r^a \le g$, we know from the proof of Theorem \ref{thm:equilibria} that $B^*(f^{\rm lo}, r^a) = B^{SE}(r^a) = 0$. When $g < r^a < c_1$, then $B(f^{\rm lo}, f^{a1}(r^a)) = B(f^{\rm lo}, f^{a2}(r^a)) = r^a - g$, therefore by the observation above, $r^{\rm lo}$ is a SE routing policy, and $B^*(f^{\rm lo}, r^a) = B^{SE}(r^a)$.

We now let $r^a > c_2$ - the other possible scenario when $\pi^a \cap [c_1, c_2] = \emptyset$. Here we denote $h$ as the value of the maximization in \eqref{eq:high_a}. When $r^a \ge C(E) - h$, we know from the proof of Theorem \ref{thm:equilibria} that $B^*(f^{\rm hi}, r^a) = B^{SE}(r^a) = r + r^a - C(E)$. When $c_1 < r^a < h$, then Theorem \ref{thm:equilibria} also informs that there must always be an edge $e$ where $B_e = 0$, in the two-link case, one edge is fully blocked and the other has no routed traffic blocked. It follows then that $B(f^{\rm hi}, f^{a1}(r^a)) = B(f^{\rm hi}, f^{a2}(r^a)) = h$, and $f^{\rm hi}$ is a SE routing policy. We conclude that when $\pi^a \cap [c_1, c_2] = \emptyset$, then $V(\pi^a) = 0$.

For the remainder of the proof, we consider the case where $\pi^a \cap [c_1, c_2]$ is nonempty. We leverage the following lemma which simplifies the expression for $B^*(f, r^a) - B^{\rm SE}(r^a)$.
\begin{lemma}
    For a two-link network, if $r^a \in [c_1, c_2]$, then for any $f$,
    \begin{equation}
        B^*(f, r^a) - B^{\rm SE}(r^a) = |f_1 - (r, + r^a - c_2)/2|
    \end{equation}
    
    \begin{proof}
        When $r^a \in [c_1, c_2]$, then we know from Lemma \ref{lem:2attacks} that for any $f$,
        \begin{align}
            B^*(f, r^a) = \max \{ &B(f, f^{a1}(r^a), B(f, f^{a2}(r^a))\} \\
            %= \max\{& f_1 + \max\{f_2 + \tilde{\gamma}_i - c_1 - c_2, 0\}, \nonumber \\
            %&\max\{f_2 + \tilde{\gamma}_i - c_2, 0\} \}, \\
            %= \max\{& f_1, r + \tilde{\gamma}_i - c_1 - c_2, f_2 + \tilde{\gamma}_i - c_2 \} \\
            = \max\{& f_1, r - f_1 + r^a - c_2 \}
        \end{align}
        From the observation made above, a SE routing policy is therefore one where $f_1 = r - f_1 + r^a - c_2$, i.e., $f$ such that 
        \begin{align}
            f_1 = (r + r^a - c_2)/2, \ 
            f_2 = (r - r^a + c_2)/2
        \end{align}
        satisfies \eqref{eq:se_r_simp}. It follows then for any $f$ that
        \begin{align}
            B^*(f, r^a) - B^{SE}(r^a) = \max\{& f_1, r - f_1 + r^a - c_2 \} \nonumber \\
            &- (r + r^a - c_2)/2, \\
            %= \max\{f_1 - (r + \tilde{\gamma}_i - c_2)/2, -f_1 + (r + \tilde{\gamma}_i - c_2)/2 \}, \\
            = |f_1 - (r + r^a - c_2)/2|.
        \end{align}
    \end{proof}
\end{lemma}

As argued in the proof of Lemma \ref{lem:parallel}, $\lopi$ need not be included in the maximization in \eqref{eq:mininp} if $\lopi \le c_1$ and $\hipi$ need not be included if $\hipi \ge c_2$. Therefore, our calculation of $V(\pi^a)$ can be further simplified:
\begin{align} \label{eq:valsimp2}
    V(\pi^a) = \min_{f \in \mathcal{F}} \max \{ & B^*(f, \lora) - B^{SE}(\lora), \nonumber \\
    &B^*(f, \hira) - B^{SE}(\hira) \}, \\
    = \min_{f \in \mathcal{F}} \max \{ & |f_1 - (r + \lora - c_2)/2| , \nonumber\\
    & |f_1 - (r + \hira - c_2)/2| \}, \label{eq:valsimp3}
\end{align}
where $\lora := \max\{c_1, \lopi\}$ and $\hira := \min\{c_2, \hipi\}$. This implies that the minimizing value of $f_1$ in \eqref{eq:valsimp3} is halfway between $(r + \lora - c_2)/2$ and $(r + \hira - c_2)/2$, i.e., 
\begin{equation}
    f_1 = (2r + \lora + \hira - 2c_2)/4,
\end{equation}
which implies that $V(\pi^a) = (\hira - \lora)/4$. $\blacksquare$

\section{Conclusion}

In this paper, we studied a particular set of network routing games, wherein the attacker has full knowledge of the router policy before choosing its own policy. We showed that choosing such a best response attack policy is an NP-Hard problem over the class of parallel networks. We showed that in such networks, a SE policy is also a NE policy when the attack either doesn't have enough attack power to affect anything, or where the attacker can block nearly everything. We concluded with a study on two-link networks and how the router's uncertainty of the attack power can affect how much traffic is blocked. We also gave a method for designing routing policies to be as robust as possible against such uncertainty.

Future work will focus on expanding this value of information study first to parallel networks, and then to the set of all networks. Another path is to understand, when the routing policy is not centralized, but distributed, how each router can be incentivized to use local information to determine the proper routing policy.

\bibliographystyle{IEEEtran}
\bibliography{util/refs}

\section*{Appendix}

\begin{appendices}

\subsection{Proof for Lemma \ref{lem:parallel}} \label{app:parallelproof}

Fix $r$ and $f \in \mathcal{F}(r)$. Since all parameters except $r^a$ are fixed, we use the notation $B^*(r^a)$ and $B^{\rm SE}(r^a)$ to emphasize that we are considering how much traffic is blocked as $r^a$ varies.

To prove this lemma, we claim the following to be true:
\begin{enumerate}
    \item $B^*(r^a)$ is a continuous function.
    \item Suppose $r^a$ is such that there exists a best response $f^a$ and $e \in E$ where $c_e - f_e \leq f^a_e < c_e$. Then there exists $\varepsilon > 0$ such that
    \begin{equation}
        \frac{B^*(r^a + \delta) - B^*(r^a)}{\delta} = 1 \text{ for all } 0 < \delta < \varepsilon. \label{eq:1der}
    \end{equation}
    Otherwise, if no such $f^a$, $e$ exist, then there is $\varepsilon > 0$ such that
    \begin{equation}
        \frac{B^*(r^a + \delta) - B^*(r^a)}{\delta} = 0 \text{ for all } 0 < \delta < \varepsilon. \label{eq:0der}
    \end{equation}
        In words, $r^a$ is the lower boundary of a neighborhood where the derivative of $B^*(r^a)$ is 1 for all points in the neighborhood or 0 for all points in the neighborhood.
    \item If there exists $\varepsilon > 0$ such that
        \begin{align}
            &\frac{B^*(r^a + \delta) - B^*(r^a)}{\delta} = 0, \text{ and } \label{eq:slope1forward}\\
            &\frac{B^*(r^a) - B^*(r^a - \delta)}{\delta} = 1, \label{eq:slopeback}
        \end{align}
        for all $0 < \delta \leq \varepsilon$, then $r^a \in \alpha$.
    \item On a plot of $B^{\rm SE}(r^a)$ vs $r^a$, the slope of the line between any two points is in the interval $[0, 1]$. \label{st:slopeinterval}
\end{enumerate}

Assuming the claims are true, claims 1 and 2 imply that $B^*(r^a)$ is a continuous piecewise linear function, where the slope of each line is either 1 or 0. By claim 4, $B^*(r^a) - B^{\rm SE}(r^a)$ is increasing when the slope of $B^*(r^a)$ is 1, and decreasing when the slope of $B^*(r^a)$ is 0. Therefore, the max of $B^*(r^a) - B^{\rm SE}(r^a)$ must occur at some value of $r^a$ where the slope of $B^*(r^a)$ changes from 1 to 0. By claim 3, all such values of $r^a$ are contained in $\alpha$. In the case where $\alpha \setminus \pi^a$ is nonempty, we include the boundary points $\lopi$ and $\hipi$ as possible values where the max on the interval $\pi^a$ can occur.

Now we prove each of the claims. First we show that $B^*(r^a)$ is continuous. Observe that when $r^a$ increases (decreases) by $\varepsilon > 0$, $B^*(r^a)$ can increase (decrease) by no more than $\varepsilon$. More formally,
\begin{equation} \label{eq:cont}
    |r^a - \hat{r}^a| < \varepsilon \implies |B^*(r^a) - B^*(\hat{r}^a)| < \varepsilon,
\end{equation}
and thus the function is continuous.

To show Claim 2, suppose that $r^a$ is such that there exists a best response attack $f^a$ where $c_e - f_e \leq f^a_e < c_e$ for some $e \in E$. Increasing $r^a$ (and $f^a_e$) by $\delta$ allows the attacker to increase $B(f, f^a)$ by $\delta$. Therefore, $B^*(r^a + \delta) = B^*(r^a) + \delta$, which implies \eqref{eq:1der}.

Now suppose that $r^a$ is such that no such $f^a, e$ exist, i.e., that for any best response attack policy $f^a$ and for all $e \in E$, either
\begin{align}
    & f^a_e = c_e \text{ or } \label{eq:r1}\\
    & f^a_e < c_e - f_e, \label{eq:r2}
\end{align}
If $\delta$ is small enough so that \eqref{eq:r2} can be replaced with $f_e^a < c_e - f_e - \delta$, then increasing $r^a$ by $\delta$ cannot increase $B^*(r^a)$. This implies \eqref{eq:0der}.

To prove claim 3, we state an implication of claim 2: if \eqref{eq:slope1forward} is satisfied, then no best response $f^a$, $e$ exist where $c_e - f_e \leq f^a_e < c_e$. However, \eqref{eq:slopeback} implies that $r^a$ is also an upper boundary of a neighborhood  where such an $f^a$ and $e$ exist. The only both statements can be true is if $f^a_e \in \{0, c_e\}$ for all $e$ and for all optimal $f^a$. This implies that $r^a = C(E')$ for some $E' \subseteq E$, i.e., that $r^a \in \pi^a$. 

We now prove claim 4. The function $B^{\rm SE}(r^a)$ must be nondecreasing, since any attack policy that can be implemented with low $r^a$ can also be carried out with high $r^a$. Equation \eqref{eq:cont} shows that the slope of the line between any two points on $B^{\rm SE}(r^a)$ is $\leq 1$, so we conclude that the claim holds. $\blacksquare$

\end{appendices}

\end{document}